# INTELLIGENT COMPUTATIONAL MODEL FOR THE CLASSIFICATION OF COVID-19 WITH CHEST RADIOGRAPHY COMPARED TO OTHER RESPIRATORY DISEASES

Paula Santos[1, 2]

[1]Department of Psychology, University of São Paulo, Ribeirão Preto, Brazil
[2]Department of Head and Neck Surgery, Ophthalmology and Otorhinolaryngology, University of São Paulo, Ribeirão Preto, Brazil

## ABSTRACT

*Lung X-ray images, if processed using statistical and computational methods, can distinguish pneumonia from COVID-19. The present work shows that it is possible to extract lung X-ray characteristics to improve the methods of examining and diagnosing patients with suspected COVID-19, distinguishing them from malaria, dengue, H1N1, tuberculosis, and Streptococcus pneumonia. More precisely, an intelligent computational model was developed to process lung X-ray images and classify whether the image is of a patient with COVID-19. The images were processed and extracted their characteristics. These characteristics were the input data for an unsupervised statistical learning method, PCA, and clustering, which identified specific attributes of X-ray images with Covid-19. The introduction of statistical models allowed a fast algorithm, which used the X-means clustering method associated with the Bayesian Information Criterion (CIB). The developed algorithm efficiently distinguished each pulmonary pathology from X-ray images. The method exhibited excellent sensitivity. The average recognition accuracy of COVID-19 was 0.93 ± 0.051.*

## 1. INTRODUCTION

In November 2019, the first cases of COVID-19 were detecting by China's health authorities. After a few weeks, the virus-infected Wuhan's city and, in March, spread globally.

In Brazil, the Ministry of Health divided the pandemic into two phases for better management: containment and mitigation. The first phase cases were attributing international travel or contact with sick people who traveled abroad. In the mitigation phase, the Ministry of Health recognized the occurrence of community transmission, from person to person, in the country - a late recognition, since there were already deaths unrelated to the transmission chains involving travelers [1].

In this context, one of the first steps towards adopting measures of social isolation and hospitalizations is to know who has been contaminated by COVID-19. The most appropriate tests for COVID-19 are molecular tests, but they can take 24 to 48 hours to be performing. In pandemic conditions, this period can last between 5 and 10 days due to many requests, lack of equipment, and health professionals' help. Therefore, rapid tests for mapping and screening patients are necessary. As X-Ray and lung tomography tests indicate patients with respiratory problems, these tests may answer this demand.

Despite several studies on the diagnosis of pneumonia regarding the criteria for confirmation and classification of pneumonia cases, many questions remain open. Thus, to avoid misinterpretation, this work was based on three underlying assumptions [2, 3]:

Pneumonia must be defined as an acute infection of the lung parenchyma by various pathogens, excluding the condition of bronchiolitis.

1. Defining pneumonia as a group of specific co-infections with different characteristics is not a line to be followed since the etiological agents' identification is not always possible.
2. Like other criteria, different types of pneumonia can be classified into more homogeneous groups, producing faster diagnosis advances.

Image recognition and analysis were revolutionizing with the introduction of deep learning, which allowed for unprecedented leaps in performance. The rapid advancement of these technologies expands the possibilities of automated, accurate, accessible, and economical medical diagnostics. Still, smart models are faster than humans and can be implemented on a large scale due to clouds' power or even at the edge. Thus, artificial intelligence techniques can help to compare and group similar types of pneumonia.

The present work proposes a criterion to classify pneumonia cases based on pulmonary radiographs. The images analyzed were of patients with COVID-19 and with common bacterial or viral pneumonia. The extraction method used was Haralick, Wavelets, and we used the Bayesian Information Criterion (CIB) as a probabilistic model, and thus we used a decision tree for the classification of images.

## 2. MATERIAL AND METHODS

### 2.1. Material Collection

We searched for articles and repositories that could indicate the signs on chest radiographs (Table 1) before comparing some characteristics present in COVID-19 with tuberculosis, H1N1, dengue, and malaria. The main characteristics highlighted were: pleural effusion, ground-glass opacity, pulmonary edema, rounded morphology of opacities, and bronchitis. Subsequently, we grouped the images into three categories: pneumonia type 1 (tuberculosis and Streptococcus pneumonia), pneumonia type 2 (malaria and dengue), and pneumonia type 3 (COVID-19). The separation was based on similar descriptions characteristic of these pathologies (Table 2) [4-25]. For this study, we used a total of 3800 chest X-ray images posteroanterior and anteroposterior positions of COVID-19 (1800), dengue (100), tuberculosis (730), Streptococcus pneumonia (200), malaria (270), normal (700). The set of images was acquired from repositories [4-25](Figure 1). Also, we use only images that a doctor has already diagnosed. That is, we use images from defined case studies. After separating the set of images, we perform manual segmentation of the pulmonary images and then look for five features that are often detectable on chest radiography was doing considering the descriptions made in case studies and articles published in the medical field to describe the intensity in the respective conditions, as shown in Table 1.

Table 1. Manifestations in the images: Diseases typical of countries like Brazil that can trigger viral and bacterial pneumonia.

| | Pleural effusion | Ground glass opacity | Pulmonary edema | Rounded morphological opacities | Bronchitis |
|---|---|---|---|---|---|
| **COVID-19** | X***(11) | X*(10) | X**(13) | X*(10) | X***(11) |
| **Dengue** | X*(12,17) | X**(12) | X**(12,17) | X***(12) | X***(12) |
| **Malaria** | X**(8) | X**(1) | X**(13) | X**(1) | X**(8) |
| ***Streptococcus pneumoniae*** | X**(9) | X**(9) | X***(6) | X**(6, 9) | X*(6, 9) |
| **Tuberculosis** | X*(15, 18) | X**(2) | X**(14) | X*(15,2,18) | X*(15, 18) |

\* ordinary; ** 40-60% of cases; *** uncommon; mon * are significantly different; ** are not significantly different.

Table 2. Image groups for character extraction and model training

| Pneumonia type 1 | Pneumonia type 2 | Pneumonia type 3 |
|---|---|---|
| Tuberculosis | Malaria | COVID-19 |
| *Streptococcus pneumoniae* | Dengue | - |

## 3. INTELLIGENT ARTIFICIAL MODEL

The model was inspired in ChestNet [26] a Neural Network for support in diagnosis for problems pulmonaries. The workflow for analysis was developed for identifying the characteristics of regions of an image, using algorithms of Wavelets and Haralick extraction attributes of the image. These texture attributes are essential because they determine partners in the lung and rub through clusterization of pixels in different Chest X-Ray regions. Subsequently, we extract the characteristics to determine the hyperparameters of backpropagation from clustering and of data mining descriptive statistics.

The probabilistic model is essential to determine each parameter in a neural network because the descriptors auxiliary the segmentation of regions and classification from the region of interest from the pattern of variation of shades of gray or color of a given region of interest. These existing partners in physicals superficially noticeable to the human eye, bringing a significant amount of information about the superficial nature, such as smoothness and roughness.

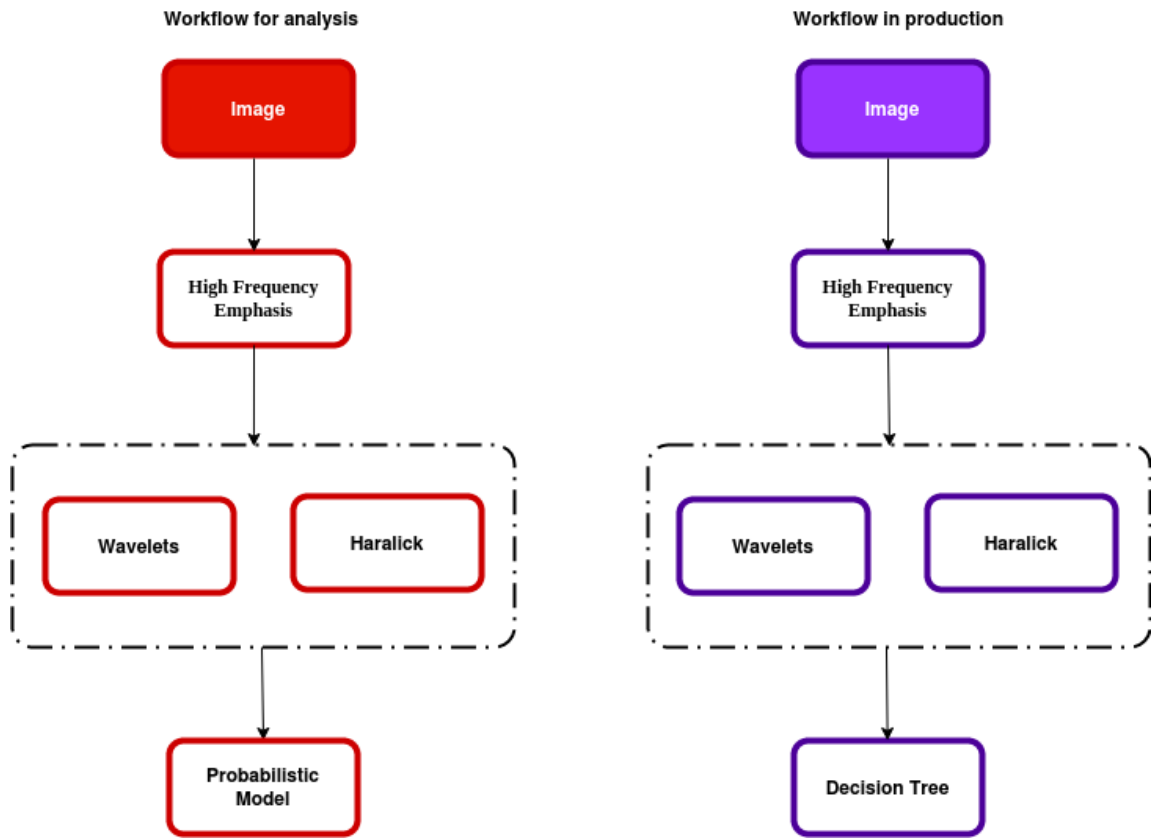

Figure 1. Workflow for analysis and production of the intelligent artificial model in Telegram.

## 3.1. Pre-Processing

This step's main objective was to improve the visualization of the bones through the implementation of the image enhancement algorithm of the High-Frequency Emphasis (HEF) filter [27]. HEF helps to sharpen an image by emphasizing the edges; since the edges usually consist of an abrupt change in the pixels' color intensity, representing the high-frequency spectrum of the image.

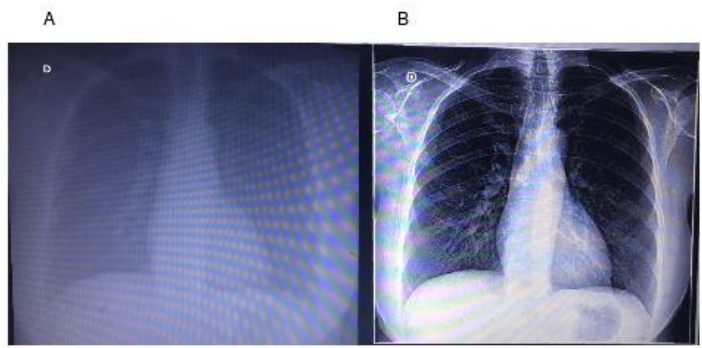

Figure 2. (A) Poor quality image; (B) After using HEF

## 3.2. Lung Segmentation Using the Mask-Regional Convolution Neural Network (Mask-RCNN).

Mask-RCNN is a deep neural network designed to solve instance segmentation problems in machine learning or computer vision [28]. For this model's training, we used 5000 lung images, segmenting the lung on the right and left sides (Figure 3).

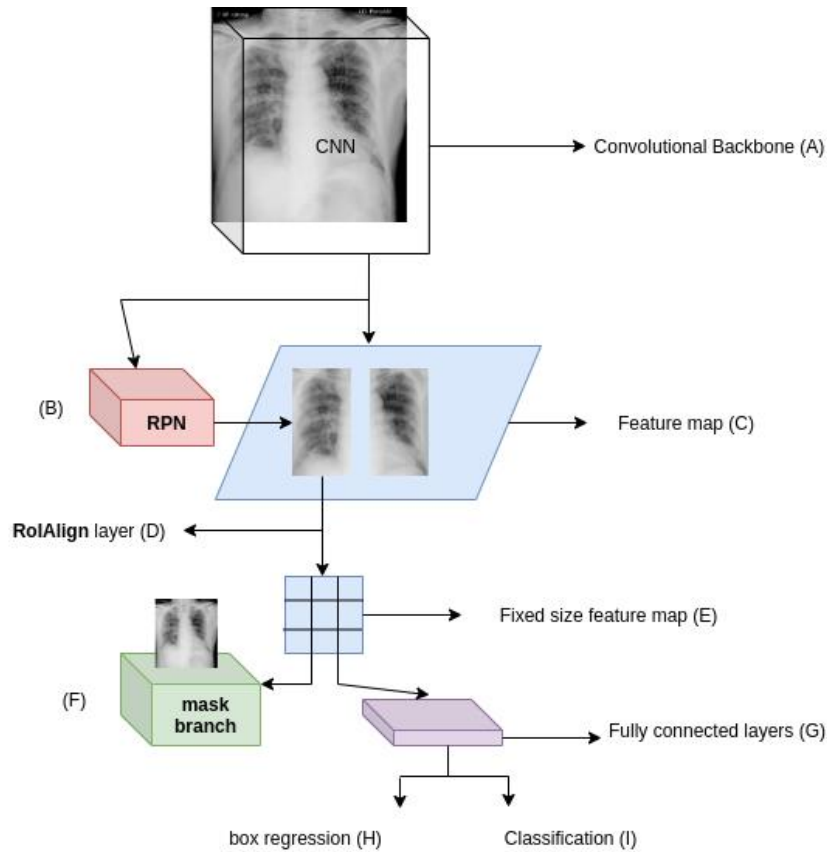

Figure 3. Mask-RCNN architecture model. (A) Convolutional Backbone of ResNet101 for extracting a map of X-ray scanned image characteristics; (B) Region Proposal Network (RPN): a neural network of small weights that provides for the bounding boxes of the lung under analysis on the characteristics map; (C) Feature map: a result of activating the output of filters applied to the image; (D) RoIAlign layer: a bilinear interpolation of nearby points on the feature map to avoid quantizing the region of interest (RoI); (E) Fixed size feature map: Reduced version extracted from the feature map; (F) Mask Branch: Mask of a fully convolutional network (RTC), which provides a segmentation lung mask for each RoI; (G) Fully connected layers use high-level RoI features by remodeling to a forecast vector; (H) Box regression: predicts the values of the pulmonary coordinates; (I) Classification for the prediction of the lung class.

## 3.3. Characteristics of X-Ray Images and Haralick Extractor

Studies show that chest radiographs are initially based on the visualization of the following three characteristics [2-3]. They are (1) Anatomical structures, such as ribs and other bones, must be visible. (2) The darker (black in the image) the color of the lungs, the more suitable is the functionality. (3) The heart and peripheral blood vessels must be visible. Using these characteristics, we applied the Haralick method (Figure 3) [29] to extract texture characteristics through their attributes, using a gray level co-occurrence matrix. The co-occurrence matrix is a square matrix whose size is the number of gray levels in the image to be analyzed. The developed algorithm calculates the distances in all possible 360 degrees and normalizes between 0 and 100. Therefore, the co-occurrence matrix contains 100 rows per 100 columns and generates by combining the distances between the current angle and their respective combinations. 10, 45, 90, and 135 degrees. After calculating this matrix, a matrix of the probability of the combinations between the gray levels was calculating. The following texture characteristics' values were calculated from this matrix: energy, entropy, variance, homogeneity, dissimilarity, and correlation measures.

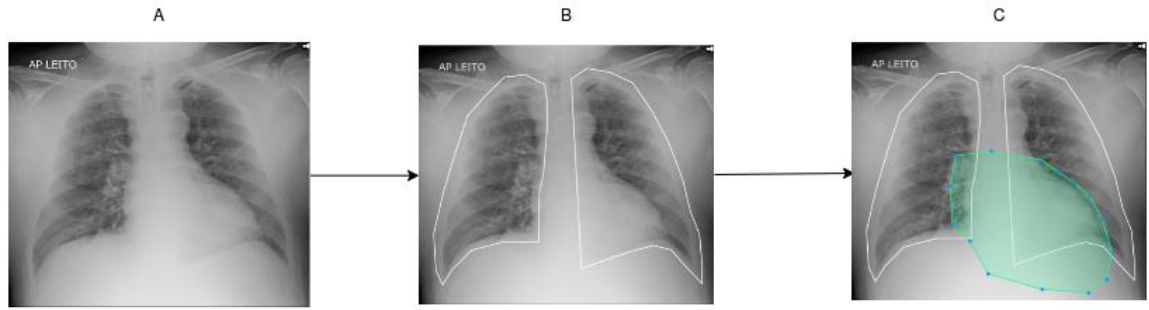

Figure 3. Regions marked for training. (A) Image of a patient with COVID-19; (B) Pulmonary marking; (C) Heart marking

### 3.4. Characteristics of X-Ray images and Wavelets

The wavelet transform can be time associated with these frequencies, making it very suitable for various fields. As an example, we can mention processing accelerometer signals for motion analysis and fault detection. Success in image compression, using the wavelet transform, is mainly attributed to innovative strategies for organizing and representing data from a transformed image. Such strategies explore implicitly or explicitly the static properties of the coefficients transformed into a wavelet pyramid. Most of the published coders recently used the pyramidal (dyadic) decomposition algorithm in the literature [33] For this project, we use the clustering of significant coefficients in a sub-band.

### 3.5. Selection of probabilistic model

The model selection problem refers to choosing the best model among a set of candidates built from combinations of parameters. Consider a sequence of models M1, M2, and Mn, with the corresponding parameters. There are many techniques for selecting the best model based on the probability ratio, and others add different types of penalty functions to the likelihood ratio. This is the case of the Akaike Information Criterion (CIA) and the Bayesian Information Criterion (CIB), both of which test two models at a time, and the two can be chosen in ascending order of the number of parameters. After that, there is a sequence of CIB and CIA values, which are optimized. This results in the number of parameters to determine which model is the best. Therefore, in the present study, we used X-means [30], an algorithm that efficiently searches the space of the clusters' locations and the number of groups to optimize the measurement of the CIB. To verify the training, testing, and validation of the model, a decision tree was used to find the hyperparameter and the inference tests [31].

### 3.6. Statistical Analysis

We used an Analysis of Variance (ANOVA) followed, when appropriate, by the Tukey-Kramer test of multiple comparisons for different sample sizes.

## 4. RESULTS AND DISCUSSION

Before analyzing the 3800 chest X-ray images, we separated the left lung from the right using Mask-RCNN, as described in the methodology, and each segmented lung pair was labeled as type 1 pneumonia, type 2 pneumonia, or type 3 pneumonia (Table 2), based on the data set information and on the literature review on signs and symptoms. The Shapiro normality test resulted in a p-value less than alpha (p-value = 4.899e-33 <0.05). Thus, the null hypothesis was rejected, and we concluded that the data were not extracted from a normal distribution. However, the information obtained from ANOVA and the Tukey-Kramer test was essential to determine the type of data distribution and the type of model. From these results, non-parametric models were used to determine the number of classes and the most appropriate classification model for this type of data (Table 3). Figure 4 shows that the characteristics were grouped according to the pathologies in pneumonia type 1, pneumonia type 2, and pneumonia type 3.

The wavelet transform can be time associated with these frequencies, making it very suitable for various fields. As an example, we can mention processing accelerometer signals for motion analysis and fault detection (Figure 5).

In this context, model selection is a problem of choosing the set of candidate models with the best performance for training data sets or estimating the model's performance using a resampling technique, such as cross-validation of k-folds. One way to use model selection involves using probabilistic statistical measures to quantify the model's performance in the training data set and the model's complexity, one of which is the Bayesian Information Criterion. The benefit of this information criterion is that it does not require a standby test, although a limitation is that they do not accept the models' uncertainty under consideration and may end up selecting straightforward models.

Table 3. Multiple comparison of means - Tukey HSD, FWER = 0.05 to assess Haralick resources.

| Comparisons | Mean difference | Significance |
|---|---|---|
| *Contrast* | | |
| Contrast 1 x Contrast 2 | -170,784 | Significant |
| Contrast 1 x Contrast 3 | -195,032 | Significant |
| Contrast 2 x Contrast 3 | -2.4248 | Significant |
| *Energy* | | |
| Energy 1 x Energy 2 | 0.031 | Significant |
| Energy 1 x Energy 3 | 42.3753 | Not significant |
| Energy 2 x Energy 3 | 423,443 | Not significant |
| *Homogeneity* | | |
| Homogeneity 1 x Homogeneity 2 | 0.031 | Significant |
| Homogeneity 1 x Homogeneity 3 | 42.3753 | Not significant |
| Homogeneity 2 x Homogeneity 3 | 423,443 | Not significant |
| *Correlação* | | |
| Correlation 1 x Correlation 2 | 0.007 | Significant |
| Correlation 1 x Correlation 3 | 101.9509 | Not significant |
| Correlation 2 x Correlation 3 | 1,019,439 | Not significant |
| *Dissimilarity* | | |
| Dissimilarity 1 x Dissimilarity 2 | 0.8779 | Significant |
| Dissimilarity 1 x Dissimilarity 3 | 532,676 | Not significant |
| Dissimilarity 2 x Dissimilarity 3 | 52.3897 | Not significant |

A                               B

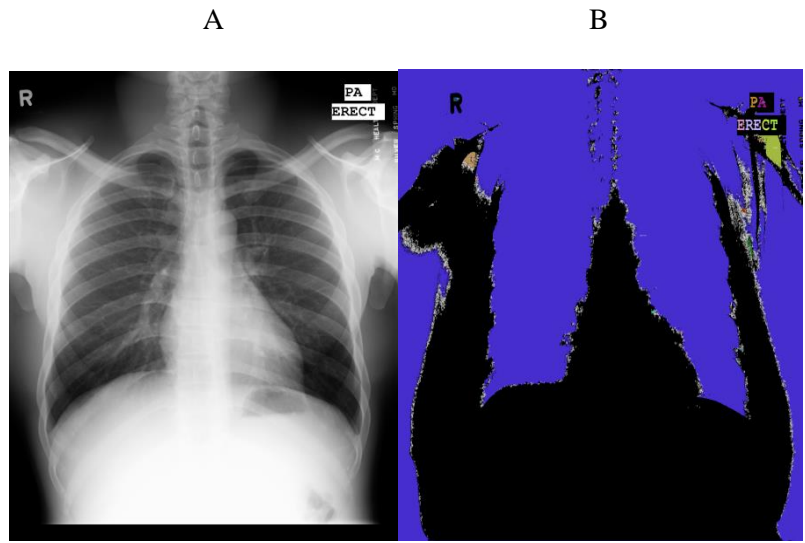

Figure 5. Algorithms of (A) Haralick and (B) Wavelets

In Figure 6, we show that the group grouping was efficient. The X-means method allows a variety of cluster K (K-means) to occur, which deals with the allocations of the clusters, repeatedly, trying to partition and maintain the resulting ideal divisions. In this segmentation, we obtained three clusters, validating the data grouping to use a decision tree model.

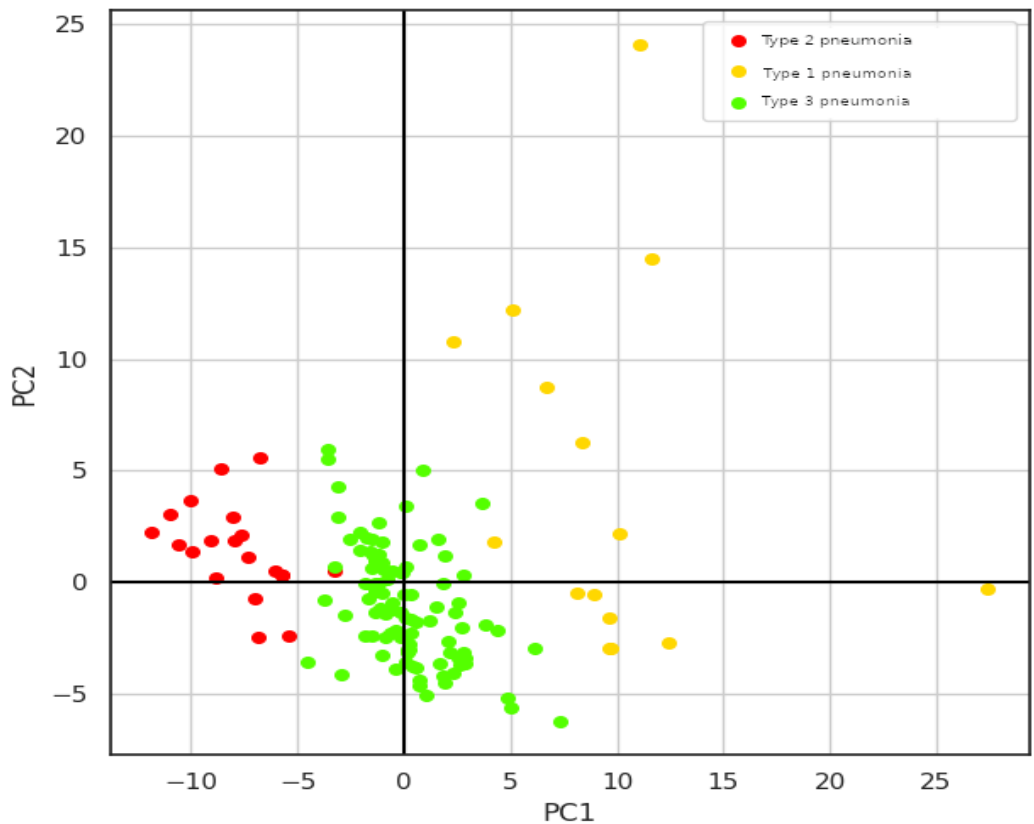

Figure 6. Scatter plot of the partition of the three clusters in the first two main components, for X-means with CIB.

The training was done with 70% of the characteristics and the test with the remaining 30%. The stratified K-Fold approach was using three scores: a minimum score of 0.95, a maximum score of 0.98, and an average score of 0.96. These results show that the average score of 0.96 presents a high assertiveness and indicates the model's high quality with real data. The optimization of the decision tree's hyperparameters was to determine the best criteria, the precision, and the standard deviation of the model (Table 4). The resulted in three scores: a minimum score of 0.95, a maximum score of 0.98, and an average score of 0.96. These results show that the average score of 0.96 is highly accurate and indicates the model's high quality with real data. The optimization of the decision tree's hyperparameters was to determine the best criteria, the precision, and the standard deviation of the model.

Table 4. Hyperparameters of the decision tree model

| Best criterion | Best maximum tree depth | Best number of components | Cross validation for model evaluation |
|---|---|---|---|
| Entropy | 12 | 3 | 0.93 ± 0.051 |

With the completion of this first stage, we carried out the test in the municipality of Itapeva. In this first stage, we tested 25 patients with suspected viral or bacterial pneumonia. We tested patients who had at least one complaint about this validation stage, such as headache, changes in taste, fever, and other complaints related to acute pneumonia. However, we tested five patients who did not meet the inclusion criteria for COVID-19 or Tuberculosis, but when the model evaluated the X-Ray, a normal patient presented COVID-19 PCR-RT test was negative, and another patient was also asymptomatic. The model was evaluated as COVID-19 and was confirmed with the PCR-RT test. In the case of the patient with Tuberculosis, he already had Tuberculosis previously, but it was positive for COVID-19 by the PCR-RT. All patients underwent the PCR-RT or Bacilloscopy test to confirm the diagnosis.

Table 5. Validation in the health unit with patients with suspected COVID-19

| | number of suspect diagnoses | number of diagnoses |
|---|---|---|
| **Normal** | 5 | 4 |
| **COVID-19** | 18 | 20 |
| **Tuberculosis** | 2 | 1 |

Currently, some studies have focused on the diagnosis by computed tomography (CT), a technique that has better sensitivity for the detection of soft tissues. However, radiography devices' availability in countries like Brazil is 1: 25,000 inhabitants, while that of CT devices is 1: 100,000 inhabitants. Some countries in Africa also offer more X-ray equipment than CT. Thus, the study of pattern recognition algorithms on chest radiographs can contribute to doctors and radiologists in diagnosing COVID-19 infection in remote regions or regions without CT devices' availability. Besides, chest X-ray diagnosis can be a screening route for isolation and/or hospitalization measures since there is a limited number of molecular test kits (the main one being RT-PCR) and, depending on the manufacturer, a high index of false-negative results may occur.

In this context, several epidemiological relevance diseases have oscillatory and periodic time patterns related to their transmission in the community. These diseases can be associated with intrinsic factors such as immunity, contact pattern, renewal, virulence rates and extrinsic factors, such as temperature, humidity, and precipitation. Among these diseases, the most common are tuberculosis, malaria, Streptococcus pneumonia, and dengue [2-7]. As these different pathologies can generate conflicting signals in diagnostic imaging, we investigated metrics that can indicate biomarkers capable of avoiding the false-positive diagnosis of COVID-19. The literature indicates that ground-glass pulmonary opacity patterns, usually with bilateral and peripheral pulmonary distribution, are emerging as a hallmark of COVID-19 infection. This disease pattern, somewhat similar to that described in previous coronavirus outbreaks, such as SARS and MERS, also fits the model that radiologists recognize as the archetypal response to acute lung injury, usually initiated by an infectious or inflammatory condition [4, 7]. Inflammation can cause ground-glass opacities in lung images, indicating consolidated dense lesions that can progressively evolve to a linear structure [8-9].

Our research efforts have shown that models using artificial intelligence can determine parameters for different groups with similar symptoms and signs. Our data mostly agree with the work of Pan et al. (2020), showing a preponderance of abnormalities in ground glass in the course of the disease. The recognition of image patterns in this group of images with similar signs and symptoms is an auxiliary tool for understanding the disease's pathophysiology since the definitive diagnosis of COVID-19 requires a positive RT-PCR test. However, current best practices recommend it as an additional test, but not for the final diagnosis of COVID-19. However, the intelligent computational model can help identify complications in the screening systems and the monitoring of pulmonary problems since there is still no effective drug for the disease, and the vaccines are still in the process of validation. The data obtained with our model suggest that the Haralick method can determine the patterns of pulmonary imaging characteristics showing pleural effusion, ground-glass opacity, pulmonary edema, rounded morphological opacities, and bronchitis. These metrics allowed the model to distinguish and significantly classify the three different pneumonia types with high accuracy. Currently, the intelligent computational model is used via Telegram by health professionals in municipalities in Minas Gerais, in Brazil (Figure 7).

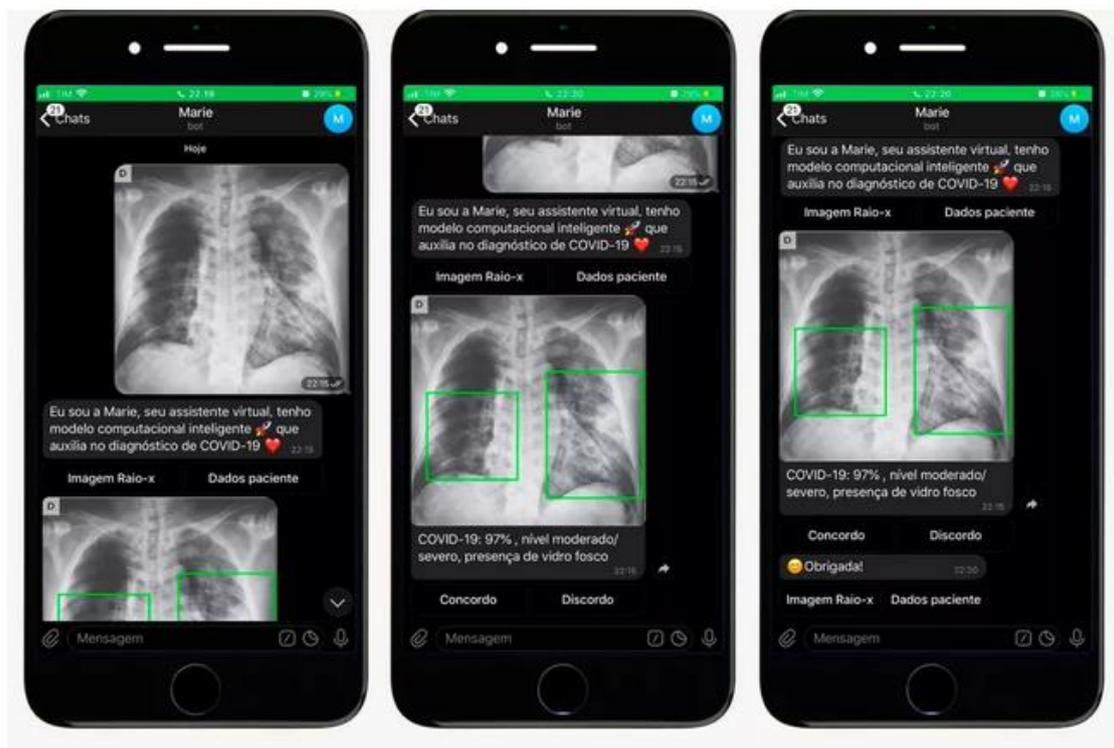

Figure 7. (A) Implementation of the intelligent computational model in the telegram for access to doctors, nurses, and radiologists.

## 5. CONCLUSIONS

Haralick's texture descriptors were useful for efficiently representing patterns of interest for image analysis and interpretation, as they showed changes in pixel intensity patterns, which were correlating with pathological changes in COVID-19. However, the new approach to extract texture characteristics by the Haralick method provided more results for the predictive analysis of pixel intensity and, when associated with unsupervised methods (X-means) and supervised methods (Mask-RCNN and Decision Tree), showed results with high accuracy. Thus, characteristics such as homogeneity, energy, dissimilarity, and correlation significantly differentiated some pathologies from the pathology of COVID-19. These results suggest that these characteristics can be used as biomarkers. These biomarkers could also be used to understand the course and stage of the disease. Our results and the preliminary test showed that chest X-rays could help healthcare professionals identify and diagnose COVID-19.

**ACKNOWLEDGMENTS**

The authors would like to thank the Ministry of Health of Brazil for making the MAIDA system's images. And the health professional Valdirene Bento and the Health Secretary of Itapeva-MG Luciano.

## AUTHOR

I obtained three bachelor's degrees in biomedical informatics and speech therapy from the University of Sao Paulo and mathematics from Anhembi Morumbi, a PhD in Sciences with a focus on Bioinformatics data in chronic pain and tinnitus models and a Post-doctorate in Psychology for the development of intelligent models for evaluation of social behaviors by the U. I also acted as Coordinator in projects of Bioinformatics and Artificial Intelligence in Medical Images in the evaluation of good or bad prognosis of childhood cancer by the Department of Pediatrics of the Hospital das Clínicas de Ribeirão Preto.

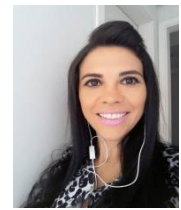